\documentclass[prb,floatfix,showpacs,twocolumn]{revtex4-1}
\usepackage[USenglish]{babel}
\usepackage{amsmath,amssymb,amsfonts}
\usepackage{epsfig}
\usepackage{subfigure}

\usepackage[colorlinks=true,linkcolor=blue,citecolor=red]{hyperref}

\newcommand{\eq}[2]
{
  \begin{equation}
    #1
    \label{#2}
  \end{equation}
}

\newcommand{\equ}[1]
{\ref{#1}}

\newcommand{\figu}[1]
{Fig.~\ref{#1}}

\def\bcen{\begin{center}}
\def\ecen{\end{center}}

             \def\d{\delta} 
\def\e{\varepsilon}          
\def\k{\kappa}              
          \def\p{\pi}          \def\s{\sigma}
           
        \def\o{\omega}

\def\CC{{\cal C}}\def\HH{{\cal H}}

\def\GG{{\cal G}}

\def\=={\equiv}

\def\qed{\raise1pt\hbox{\vrule height5pt width5pt depth0pt}}

\def\cG0{{\cal G}_0} 
\def\cG{{\cal G}}

\def\up{\uparrow}  \def\dw{\downarrow}
\def\bra{\langle} \def\ket{\rangle}

\def\ka{{\mathbf k}}   
\def\Ak{{\mathbf A}(t)}
\def\Ek{{\mathbf E}}

 \def\=={\equiv}

 \def\ep0{\epsilon_p} \def\ed0{\epsilon_d}

\begin{document}
\title{Non-equilibrium dynamics of the driven Hubbard model.}

\author{A.~Amaricci$^{1}$, C.~Weber$^{2}$, M.~Capone$^{1,3}$, G.~Kotliar$^{4}$}

\affiliation{$^1$ CNR-IOM, SISSA, Via Bonomea 265, 34136 Trieste, Italy.}
\affiliation{$^2$ Cavendish Laboratory, Cambridge University,  J.J. Thomson Ave., 
Cambridge, UK}
\affiliation{$^3$ Physics Department, University ``Sapienza'', Piazzale A.~Moro 2, 00185
Rome, Italy.}
\affiliation{$^4$ Department of Physics and Astronomy, Rutgers University, Piscataway, 
New Jersey 08854, USA} 

\date{\today}

\begin{abstract}
We investigate the dynamics of the Hubbard model in a static electric field in order to identify the conditions to reach a non-equilibrium stationary state. We show that, for a generic electric field, the convergence to a stationary state requires the coupling to a thermostating bath absorbing the work done by the external field.  Following the real-time dynamics of the system, we show that a non-equilibrium stationary state is reached for essentially any value of the coupling to the bath. We characterize the properties of such non-equilibrium stationary states by studying suitable physical observables, pointing out the existence of an analog of the Pomeranchuk effect as a function of the electric field. We map out a phase diagram in terms of dissipation and electric field strengths and identify the dissipation values in which steady current is largest for a given field.
\end{abstract}


\maketitle

\section{Introduction}
The theoretical investigation of out-of-equilibrium strongly correlated quantum systems recently generated a tremendous interest, stimulated by the development of novel experimental techniques, which allow to explore transport properties of correlated materials in a non-equilibrium regime,\cite{Giannetti09,Giannetti11,Fausti11} notwithstanding the recent achievements in the field of optically trapped cold-atoms. These experimental advances challenge the theory to develop suitable new methods to describe how a strongly correlated materials evolves when an external field pushes it out-of-equilibrium.

Dynamical mean-field theory (DMFT) is an established method to investigate correlated materials at equilibrium\cite{rmp}. The recent extension of the DMFT out of the equilibrium \cite{Freericks06,SchmidtMonien}, provides us with a reliable tool to clarify how correlation effects influence the non-equilibrium dynamics of quantum systems. Non-equilibrium DMFT has been successfully applied to study quantum quenches - sudden changes of some control parameter\cite{EcksteinKollarWerner,Eckstein10}, and to investigate driven correlated systems\cite{Freericks06,Joura08}. In this context, DMFT has been used to show how interactions favor the
formation of stationary states by suppressing Bloch oscillations of the current\cite{Freericks06,Freericks08,Freericks2,Eckstein11} and to analyze the dielectric breakdown of Mott insulators triggered by the application of large fields\cite{Oka03,Oka05,Eckstein10-1}.

In this work we focus on the role of the coupling to an external thermostat in the out-of-equilibrium dynamics of a correlated system and we show when and how it leads to a non-equilibrium stationary state (NSS). Despite its importance, the role of dissipation in driven correlated systems has only been
discussed assuming the existence of a NSS\cite{Tsuji09}, while we are not aware of any study following the real-time dynamics leading to such state. 

In the classical framework, dissipation is usually introduced by coupling the system to a
set of reservoirs that ultimately impose a suitable constraint on the equations of
motion\cite{Gallavotti99,Amaricci07}. Nevertheless, the direct extension of this approach
to the quantum regime is not straightforward because of the Hamiltonian nature of the
quantum equations\cite{Gallavotti07}. By exploiting the local nature of the DMFT, we demonstrate that a thermostatting mechanism can be realized through a non momentum-conserving coupling of the correlated electrons to a set of local fermionic baths.  

In the following, we consider the two-dimensional Hubbard model, which is the paradigm of strong correlations and minimal model to capture some important properties of the high-$\mathrm{T_c}$ superconductors. We follow the real-time dynamics when the system is driven out of equilibrium by a static electric field. We show that, for a given value of the electric field, coupling to the thermostat is necessary to reach a physically relevant NSS.
Remarkably, a physically sound NSS with finite current can be reached for almost any value of the dissipation, regardless the initial conditions. The coupling to a bath is therefore also an essentially
sufficient condition to dynamically approach the NSS. We characterize the dynamical formation of the NSS  and its properties by means of suitable physical observables. We point out the existence of an analog of the Pomeranchuk effect in the NSS as a function of the electric field. We summarize our results in a phase-diagram in terms of the coupling to the thermostat and the electric field strength.

The remainder of the paper is organized as follows: Sec. II and III respectively introduce our model and the non-equilibrium Dynamical Mean-Field Theory. Sec. IV is devoted to the results of the real-time dynamics of the driven Hubbard model, while Sec. V is focused on the characterization of the non-equilibrium stationary state. Finally, Sec. VI presents concluding remarks.


\section{Model}
We consider the two-dimensional one-band Hubbard model on a square lattice with spacing $a$.
The model Hamiltonian reads ($c=1$): 
\eq{
\HH_{c} = \sum_{\ka\s} \e(\ka - e\Ak)c_{\ka\s}^\dagger c_{\ka\s} + U\sum_{i}
{n}_{i,\up}{n}_{i,\dw}
}{model}

This model describes electrons with only nearest neighbor hopping and subject to a local  Coulomb repulsion of strength $U$. The hopping amplitude $J=1$ sets the the energy unit and determines the dispersion $\varepsilon(\mathbf{k}) = -2J[\cos(k_xa)+\cos(k_ya)]$. The system is
driven out-of-equilibrium by coupling to a constant and homogeneous electric field $\Ek$, derived from a purely vector potential $\Ak=-r(t)t\Ek $, with $\Ek = E\mathbf{Q}$ and $\mathbf{Q}=(\pi,\pi)$. The function $r(t)$ specifies the switching protocol of the field. In this work we shall consider $r(t)=\theta(t-t_0)$, corresponding to a sudden switch on of the
electric field. Different choices for $r(t)$ will be explicitly stated. The electric field unit is $eEa$. The coupling to electric field is realized via the Peierls substitution: $\ka\rightarrow\k=\ka -e\Ak$. We will work at half-filling, when the number of electrons equals that of the lattice sites, and we will only consider paramagnetic solutions without magnetic symmetry breaking.

In a solid state system the work done by the electric field on the electrons is constantly transformed into heat by various scattering mechanisms. On the other hand, the presence of some dissipation mechanism usually allows the system to reach a NSS, characterized by the flow of finite electric current. In this respect, the presence of an external thermostat is a crucial feature to maintain the internal energy conserved on average in a non-equilibrium system. 
To take into account this effect in the treatment of the problem we shall include a simple thermostating mechanism. This can be realized by locally coupling the conduction electrons to a bath of non-interacting electrons at a fixed temperature, which we set at $T=0.01$ throughout the rest of the paper. 
The coupling to the external thermostat is required to break momentum conservation. This allows the conduction electrons’ momenta, accelerated by the application of an external field, to relax and eventually drive the formation of a NSS.
 
In the rest of this work we consider the following expression for the coupling:
$$
\sum_{i\s}\sum_l V_{i l}\left(c_{i\s}^\dagger b_{il\s} + h.c.\right)
$$

For sake of simplicity, and without loss of generality, we choose the amplitude $V_{il}\== V$ to be constant and homogeneous. From now on, we also drop any reference to spin index, as we are interested in the paramagnetic phase of the model. The external thermostats are assumed to be unaffected by the presence of the electric field. In addition, the particular details of internal structure of the thermostating reservoirs are expected to be irrelevant with respect to the physics of the NSS. Thus, we consider a set of identical systems with a constant density of states with bandwidth $W$.


\section{DMFT equations}
The Kadanoff-Baym-Keldysh\cite{KadanoffBaym,Keldysh64,Danielewicz84}
formalism provides a natural framework to investigate the real-time
dynamics out of equilibrium.
In the presence of sizeable electric fields the non-equilibrium dynamics is expected to be driven by the field and by the coupling to the thermostat. Being interested in the relaxation towards the NSS, we can simplify the treatment by considering only the real-time branches in the Keldysh contour and neglecting the imaginary-time segment. As a consequence, at time $t_0$ we will simultaneously switch-on the electric field $E$ and the
interaction $U$, introducing also a correlation quench. This is not expected to strongly influence the results in the presence of the thermostat, which is able to dissipate the extra energy involved in the quench as well as the energy pumped by the field \footnote{The interaction quench is instead expected to influence the dynamics when the coupling to the bath vanishes.}.

The thermostated system obeys the following Dyson equation on the Keldysh contour $\CC$\cite{Joura08}
\eq{
G_\k(t,t')=\GG_{0\k}(t,t') + [\GG_{0\k}\cdot \Sigma_\k \cdot G_\k](t,t')
}{eq1}
where all quantities represent continuous operators of two time variables $(t,t')\in\CC$,  the symbol $\cdot$ denotes the convolution product 
$$
\left[f\cdot g\right](t,t')=\int_\CC dz f(t,z)g(z,t')
$$ 
and $\Sigma_\k(t,t')$ denotes the Keldysh self-energy function. The Eq.\equ{eq1} is expressed in terms of the ``renormalized" non-interacting lattice Green's function $\GG_{0\k}$:
\eq{
\GG_{0\k}^{-1}(t,t')=\left[G^{-1}_{0\k}(t,t') - \Sigma_{bath}(t-t')\right]
}{eq2a}
which is obtained from the ``bare" non-interacting lattice Green's function:
\eq{
G_{0\k}^{-1}(t,t') = [i\overrightarrow{\partial_t} -\e(\k)]\cdot\d_\CC(t,t')
}{eq2b}
by integrating out locally the electronic degrees of freedom of the external thermostat.
The symbol $\d_\CC$ indicates the delta function on the contour $\CC$. 
Thus, the effects of the thermostat on the non-equilibrium dynamics are taken into account by mean of an additional self-energy $\Sigma_{bath}$, which reads:
$$
\Sigma_{bath}(t-t')= V^2g(t-t')
$$
where $g(t-t')$ is the Fourier transform of the non-interacting local
bath Green's function corresponding to a flat density of states:
\eq{
g(\o)=\frac{1}{W}[\ln{\left|(W/2+\o)/(W/2 -\o)\right|} -i\pi\theta(\frac{W}{2}-|\o|)]
}{eqbath}

The bath self-energy is by construction time-translation invariant, as
the electric field does not act directly on the thermostat, whose role
is only to absorb the excess energy pumped in by the
field. Eq. (\equ{eqbath}) also identifies 
$\Lambda= V^2/W$ as an effective coupling of the electrons with the
thermal reservoirs.

Applying $\GG_{0\k}^{-1}$ on both sides of the Eq. (\equ{eq1}) and
using Eqs. (\equ{eq2a}) and (\equ{eq2b}), we can recast the Dyson equation into the Kadanoff-Baym equation, that is the equation of motion
for the Keldysh Green's function, which reads\cite{langreth}:
\eq{
\begin{split}
 &[i\overrightarrow{\partial}_t-\e(\k)]\cdot G^>_\k = I_\k^> + I_0^>\\
 &G^<_\k\cdot[-i\overleftarrow{\partial}_{t'}-\e(\k)]= I_\k^< + I_0^<\\
\end{split}
}{eqKB2}
with
\eq{
\begin{split}
I_\k^> &=\Sigma^R_\k\cdot G^>_\k +\Sigma^>_\k\cdot G^{A}_\k\,,\quad I_0^>=\Sigma_0\cdot G^>_\k\\
I_\k^< &=G_\k^R\cdot\Sigma^<_\k  + G_\k^<\cdot \Sigma^A_\k,\quad I_0^<=G_\k^<\cdot\Sigma_0 \\,
\end{split}
}{eqIk}
where $G^>$, $G^<$ and $G^R$ are, respectively the greater, lesser and retarded component of the matrix Green's functions in the Keldysh formalism, and analogous notations are used for the self-energy.

As an effect of the symmetries relating the Keldysh components, Eqs. \equ{eqKB2} determine a system of coupled ordinary first-order differential equations in the $(t,t')$-plane for the lattice Keldysh Green's function $G_\k$.\cite{HSKohler99} 
The numerical solution of this system is constructed by evolving separately the Keldysh components of the non-equilibrium Green's function in some region of the discretized $(t,t')$-plane\cite{vanLeeuwen2009}. This method permits to obtain $G_\k(t,t')$ up to arbitrary large times, once the initial conditions are specified. 
The two parameters controlling the solution of the differential equations, namely the time-step $\delta t$ and the step numbers $N_t$, have to be adjusted to the particular regime of interest. While a small value of the time-step guarantees a higher accuracy in the description of the non-equilibrium dynamics, a large step number leads to an error accumulation, ultimately ending in a breakdown the numerical solution. In our calculations we found that a time-step $\delta t=0.1$ and a number of steps $N_t=250$ permit to  accurately solve the Kadanoff-Baym equation for a generic value of the electric field. Nevertheless, a larger number of steps ($N_t\simeq1000$) or a smaller time-step ($\delta t\simeq0.01$), may become necessary to describe the non-equilibrium dynamics in presence of a small decay rate or of large electric fields.

The initial conditions for the system \equ{eqKB2} are given specifying the value of momentum distribution at an (arbitrary) initial time $t=t_0$.
In this work we set $n_\ka(t_0=0)=-iG_\ka^<(0,0)=f(\epsilon(\ka))$, with $f$ the Fermi-Dirac distribution, except where otherwise stated. 
%
\begin{figure}
\centering
 \includegraphics[width=0.4\textwidth]{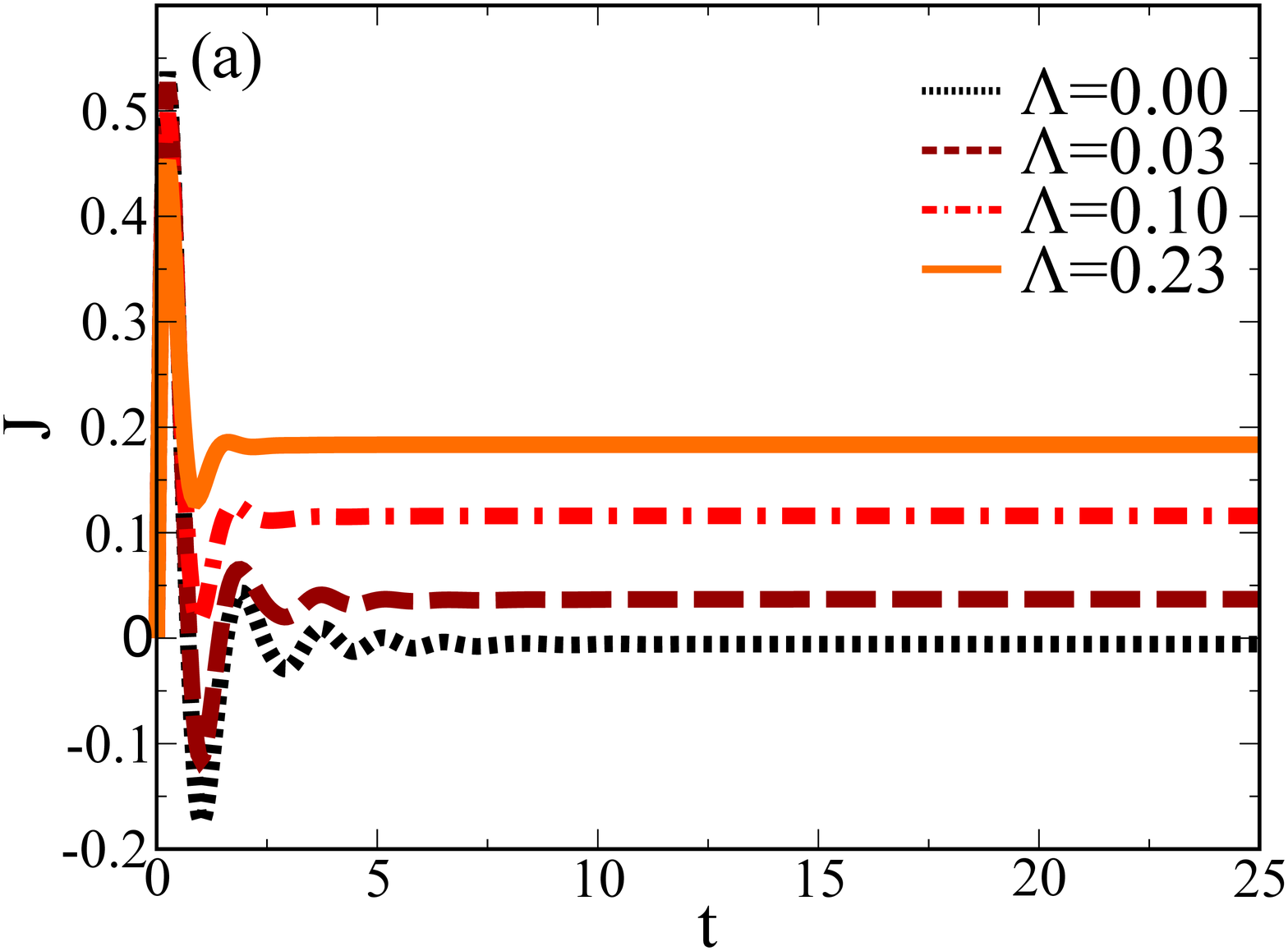}
 \includegraphics[width=0.4\textwidth]{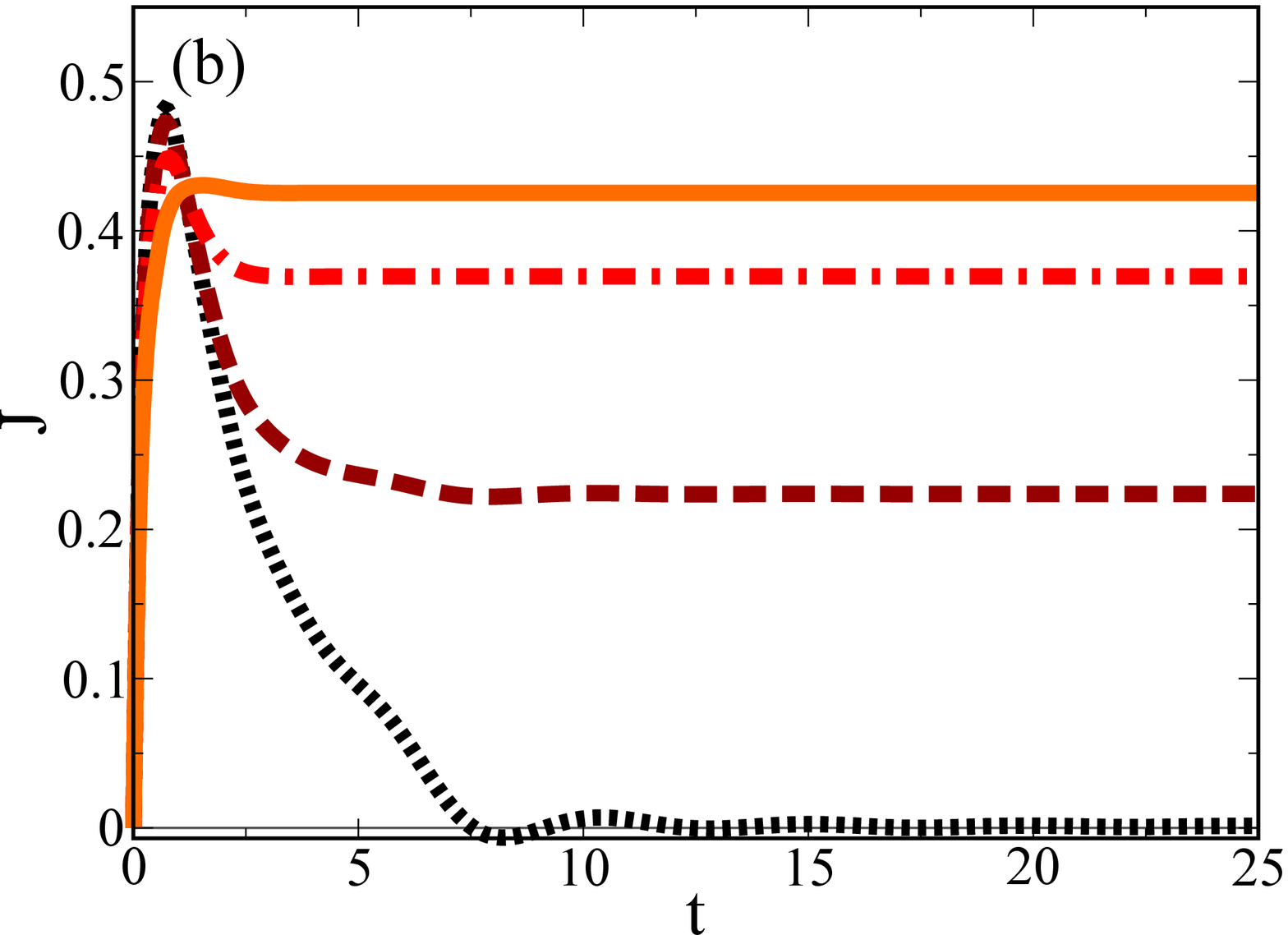}
\caption{(Color online) Dynamics of the local current $\mathrm{J}$ for $U=6.0$, increasing coupling to thermostat $\Lambda$ and $E=4.7$ (a), $E=1.26$ (b).
}
\label{fig1}
\end{figure}
%

We use the DMFT to deal with correlations in a non-perturbative way. Within DMFT the lattice self-energy is approximated by its local component, so that the momentum dependence is entirely determined by the non-interacting dispersion. The self-energy can be derived from the self-consistent solution of an impurity problem, written in terms of a local Weiss field $\GG_0(t,t')$.\cite{Freericks06} This latter describes the effective non-equilibrium medium coupled to the impurity and has to be self-consistently determined solving the following equations:
\eq{
\begin{split}
\GG_0^R &= \Gamma\cdot G^R_\mathrm{loc}\\
\GG_0^{\gtrless} &=\Gamma\cdot G_\mathrm{loc}^{\gtrless}\cdot{\Gamma^\dagger} +
\GG_0^R\cdot\Sigma^{\gtrless}\cdot{\GG_0^R}^\dagger
\end{split}
}{sc1}
with:
$$
\Gamma  = [\delta_\CC + G_\mathrm{loc}^R\cdot \Sigma^R]^{-1}
$$
The local components of the Green's functions are obtained by integrating the solution of Eqs.~(\equ{eqKB2}) over the full Brillouin zone. To close the DMFT equations it is necessary to determine the impurity self-energy $\Sigma(t,t')$. In this work we use iterated second-order perturbation theory in $U$\cite{rmp}:
\eq{
\Sigma^{\gtrless}(t,t') = U^2 [\GG_0^{\gtrless}(t,t')]^2
\GG_0^{\lessgtr}(t',t)
}{Sipt}
Equations (\ref{eqKB2}), (\ref{sc1}) and (\ref{Sipt}) define a
complete set of non-equilibrium DMFT equations. In the actual
implementation, these equations are iteratively solved until a self-consistent solution is obtained.


\section{Non-equilibrium dynamics}
The approach to the stationary non-equilibrium state can be characterized following the real-time dynamics of suitable observables, such as the local current $\mathbf{J}(t)=-ie/\p \sum_\k {\mathbf v}_\k G^<_\k(t,t)$, where ${\mathbf v}_\k=\nabla_\k \e(\k)$ is the electronic velocity. We focus on the correlated metallic phase. Our results for the local current are presented in \figu{fig1}.
The application of a constant electric field on a periodic lattice structure produces an
oscillating current (Bloch oscillations) of frequency $\omega_B=eEa$. In the absence of coupling to the external bath ($\Lambda=0$), the electron-electron interactions suppress the Bloch oscillations leading to an exponentially decaying current which eventually converges to zero at very long time [see \figu{fig1}]\cite{Freericks08,Eckstein11}. However, the relaxation
process shows two different regimes depending on the value of the interaction $U$ as found in Ref.~\onlinecite{Eckstein11}. 
Indeed, for weak interactions the dynamics is characterized by the presence of current oscillations, which suddenly disappear in the strong interaction regime where a simple exponential decaying dynamics sets in.

\begin{figure}
 \includegraphics[width=0.234\textwidth]{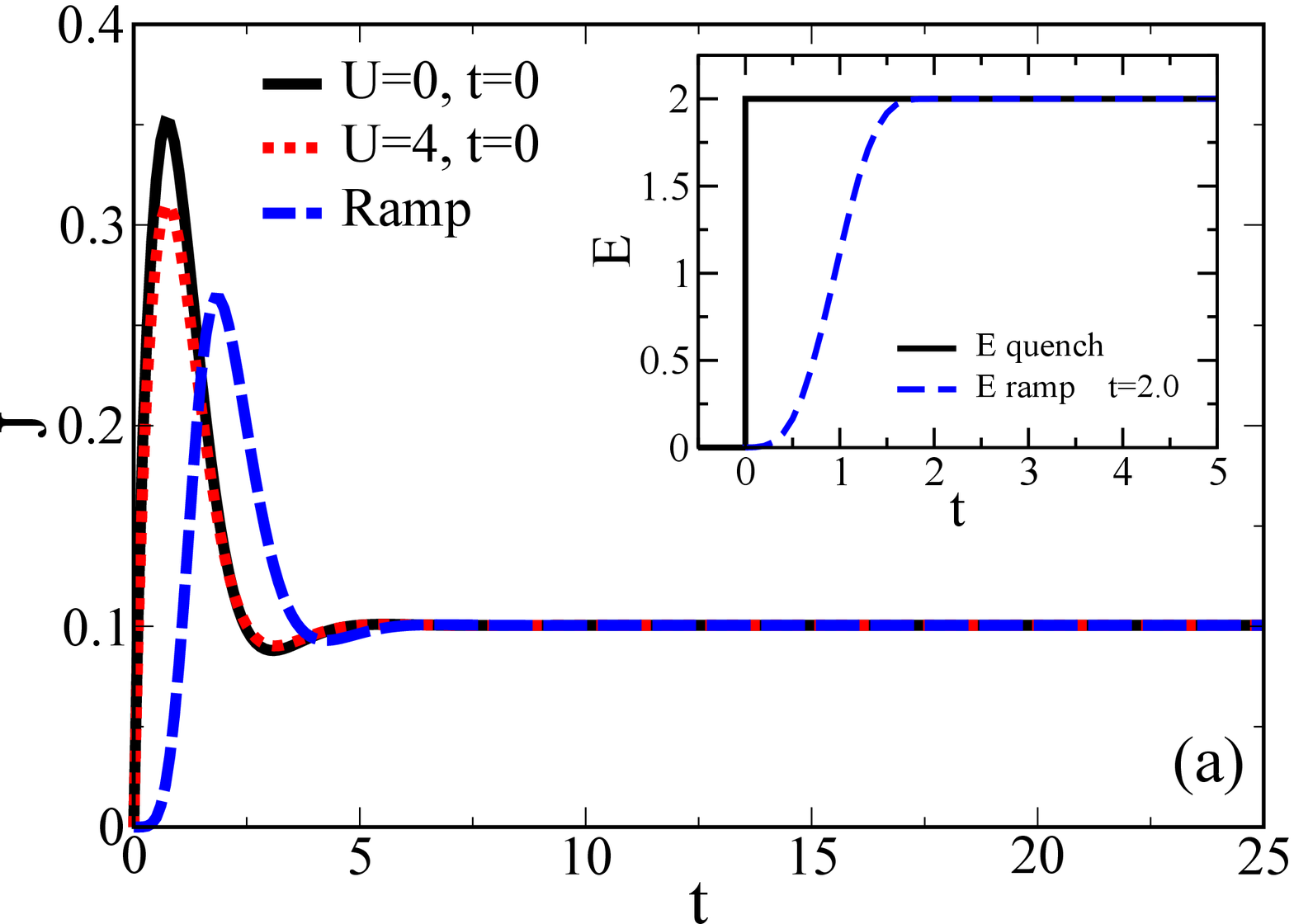}
 \includegraphics[width=0.23\textwidth]{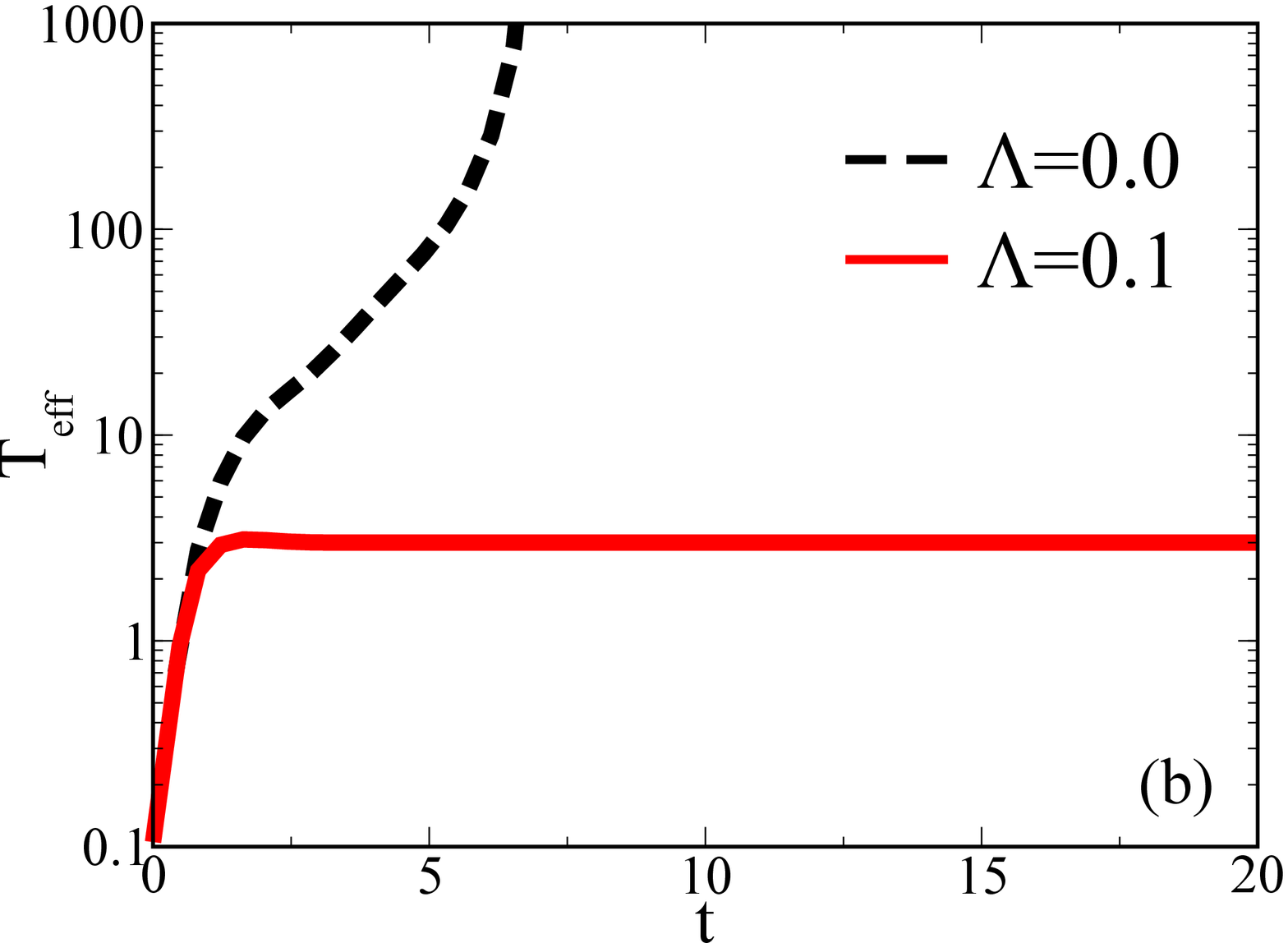}
\caption{(Color online)
(a) dynamics of the local current from different realizations of the initial conditions: electric field quench on the non-interacting $U=0.0$ (black) and interacting $U=4.0$ state (red), smooth switch-on of the electric field for $U=0.0$ and $\tau=2.0$ (blue, cf. text). The data are for $E=2.0$ and $\Lambda=0.05$. The inset shows the profile of the electric field.
(b) Effective temperature $T_\mathrm{eff}(\Omega)$ as a function of time for $U=6$, $E=1.9$. 
}
\label{fig2}
\end{figure}

The non-equilibrium dynamics of the system dramatically changes when
the system is coupled to the thermostat $(\Lambda>0$). The local current relaxes to a finite value corresponding to the formation of a NSS. This effect is detailed in \figu{fig1} for two different values of the electric field and $U=6$. 
The relaxation dynamics leading to the NSS is characterized by an increased damping of the Bloch oscillations, due to the increased scattering of the conduction electrons provided by the coupling with the thermostat. The relaxation time required to reach the NSS decreases with increasing coupling to the external bath, so that for the largest investigated coupling only few time units are required for the system to relax.

The approach to the NSS is found to be independent from the initial conditions, confirming that the non-equilibrium physics is governed by the field and the dissipation term. In \figu{fig2}(a) we compare the solutions obtained using initial conditions $r(t)=\theta(t)$ and $n_\ka(0)=f(\e(\ka))$ with the solutions obtained: (i) starting from the interacting equilibrium momentum distribution for $U=4$, (ii) using smooth switching of the electric field $r(t)=[1 - 3/2\cos{(\pi t/\tau)} + 1/2\cos{(\pi t/\tau)}^3]/2$.
The convergence to the same NSS is evident, validating the simplifying approximations discussed above.

\begin{figure}
\centering
 \includegraphics[width=0.4\textwidth]{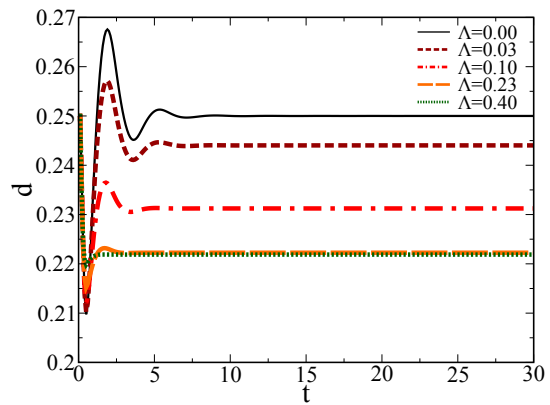}
\caption{(Color online) Non-equilibrium dynamics of the double occupations $d$ for $U=6$, $E=1.25$ and increasing $\Lambda$.}
\label{fig3}
\end{figure}

A better understanding of the effects introduced by the inclusion of a thermostat on the non-equilibrium dynamics is obtained by looking at the time-evolution of the effective temperature\cite{Eckstein11,Mitra06,Oka10}. This is defined, at any time, as the
temperature $T_{\mathrm{eff}}$ associated to an equilibrium solution with the same energy $\Omega(t)=\bra K\ket+\bra V\ket$ of the
non-equilibrium state and the same value of the interaction $T_\mathrm{eff}:\,\Omega(t)\stackrel{!}=\bra e^{-\HH/T_\mathrm{eff}} \HH\ket$.

For $\Lambda=0$ the effective temperature rapidly diverges as a function of time  because the system is unable to dissipate the energy constantly injected by the electric field. In the presence of dissipation the effective temperature relaxes instead to a finite value, determined by the balance between the action of the forcing field and the dissipation of high-energy electrons. The coupling to the thermostat provides indeed  a channel which  prevents the population of states with high energy and momentum and
allows the conservation (on average) of the total energy and the approach to a constant effective temperature.

Finally, we investigated the double occupancy $d$, a key quantity characterizing the correlated systems which is proportional to the potential energy and measures the effectiveness of correlations. The real-time evolution of double occupancy provides useful information about the dynamics of the electronic excitations in the systems. In \figu{fig3} we illustrate the formation of the NSS from the dynamics of $d$. In the non-thermostated case ($\Lambda=0$) $d$ relaxes to the non-interacting value $d=1/4$, corresponding to the excitations of all the available electronic degrees of freedom and in agreement with the already discussed divergence of the effective temperature. 
Conversely, a finite coupling to the thermostat induces the relaxation to a NSS, which is characterized by a smaller value of the double occupancy $d<1/4$. As observed for the electric current, the relaxation time to the stationary state is gradually reduced upon increasing the coupling to the thermostat.

\begin{figure}
\centering
 \includegraphics[width=0.23\textwidth]{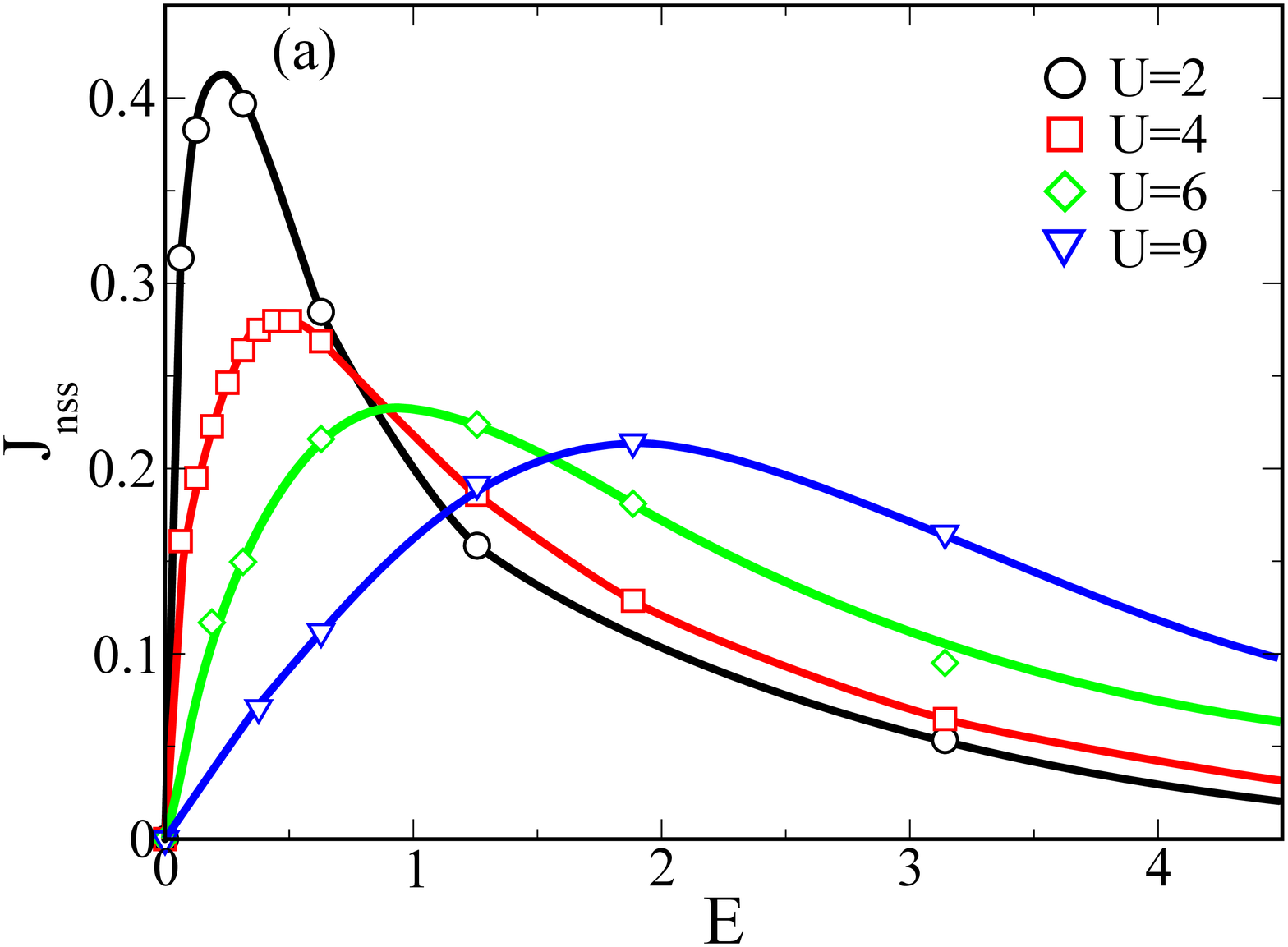}
 \includegraphics[width=0.23\textwidth]{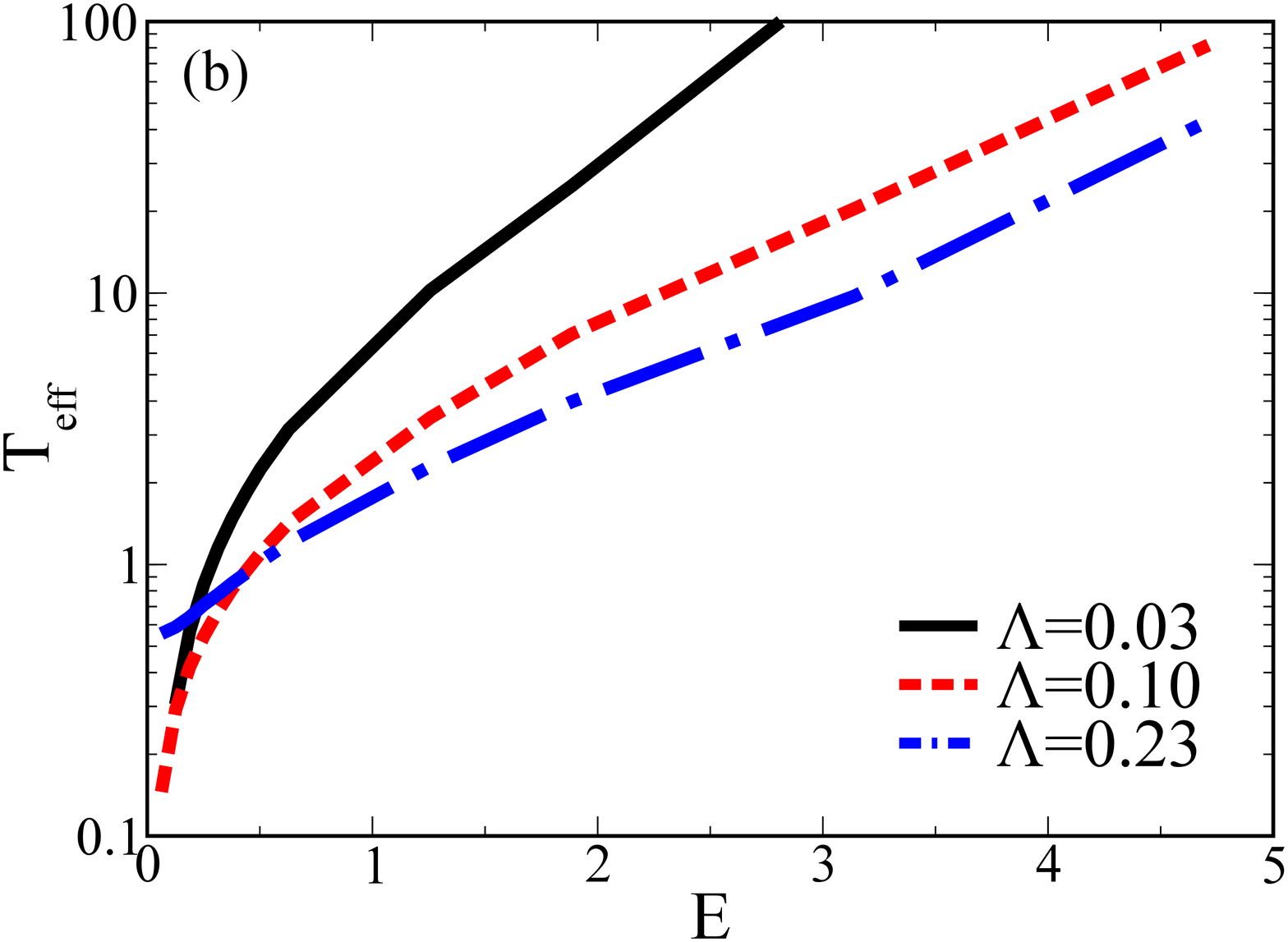}
\caption{(Color online) (a) linear/non-linear crossover of the stationary current as a function of the electric field $E$. Data for $\Lambda=0.025$ and increasing $U$. (b) Effective temperature $T_\mathrm{eff}(\Omega)$ of the NSS as a function of the applied external field and for increasing coupling to the thermostat. Data are for $U=6$ and $E=1.9$.}
\label{fig4}
\end{figure}


\section{Stationary states}
Having illustrated the formation of the NSS through the non-equilibrium dynamics of the systems, we now turn our attention to the physical properties which characterize the NSS in presence of the thermostat. 
To begin with, we plot in \figu{fig4}(a) the steady current $\mathrm{J}_\mathrm{nss}$, which shows a linear/nonlinear crossover as a function of the electric field, as repeatedly reported in similar problems \cite{Mierzejewski11,Aron11}. At small fields the current is linear in the field $E$, as expected by continuity with perturbed equilibrium state, then it reaches a maximum before decreasing as the field is further increased. The existence of a residual current at large values of the field is an effect due to the presence of the thermostat, which allows conduction electrons to bridge between Wannier-Stark states\cite{Joura08}.

In the right panel of \figu{fig4} we show the behavior of the effective temperature of the NSS as a function of the electric field $E$. For any given coupling $\Lambda$, $T_\mathrm{eff}$ is a monotonically increasing function of the field so that arbitrarily large fields imply arbitrarily large $T_\mathrm{eff}$. For large fields the effective temperature is naturally reduced increasing
the coupling $\Lambda$. 

On the other hand, for small $E$, the effective temperature $T_\mathrm{eff}$ increases with $\Lambda$. This behavior can be easily understood. For a finite electric field, a small coupling with the bath is sufficient to compensate the effects of the external field, leading to the formation of a NSS with a finite effective temperature $T_\mathrm{eff}$.
Upon reducing the field $E$ at fixed $\Lambda$, we eventually reach a small-field region in which the effect of the thermostat is not balanced by the electric field, leading to the excitation of the high-energy electronic states and ultimately to a non generic value of the low-field effective temperature. This effect is more pronounced for larger $\Lambda$, which leads to the observed larger $T_\mathrm{eff}$ for larger values of the coupling.

\begin{figure}
\centering
\includegraphics[width=0.4\textwidth]{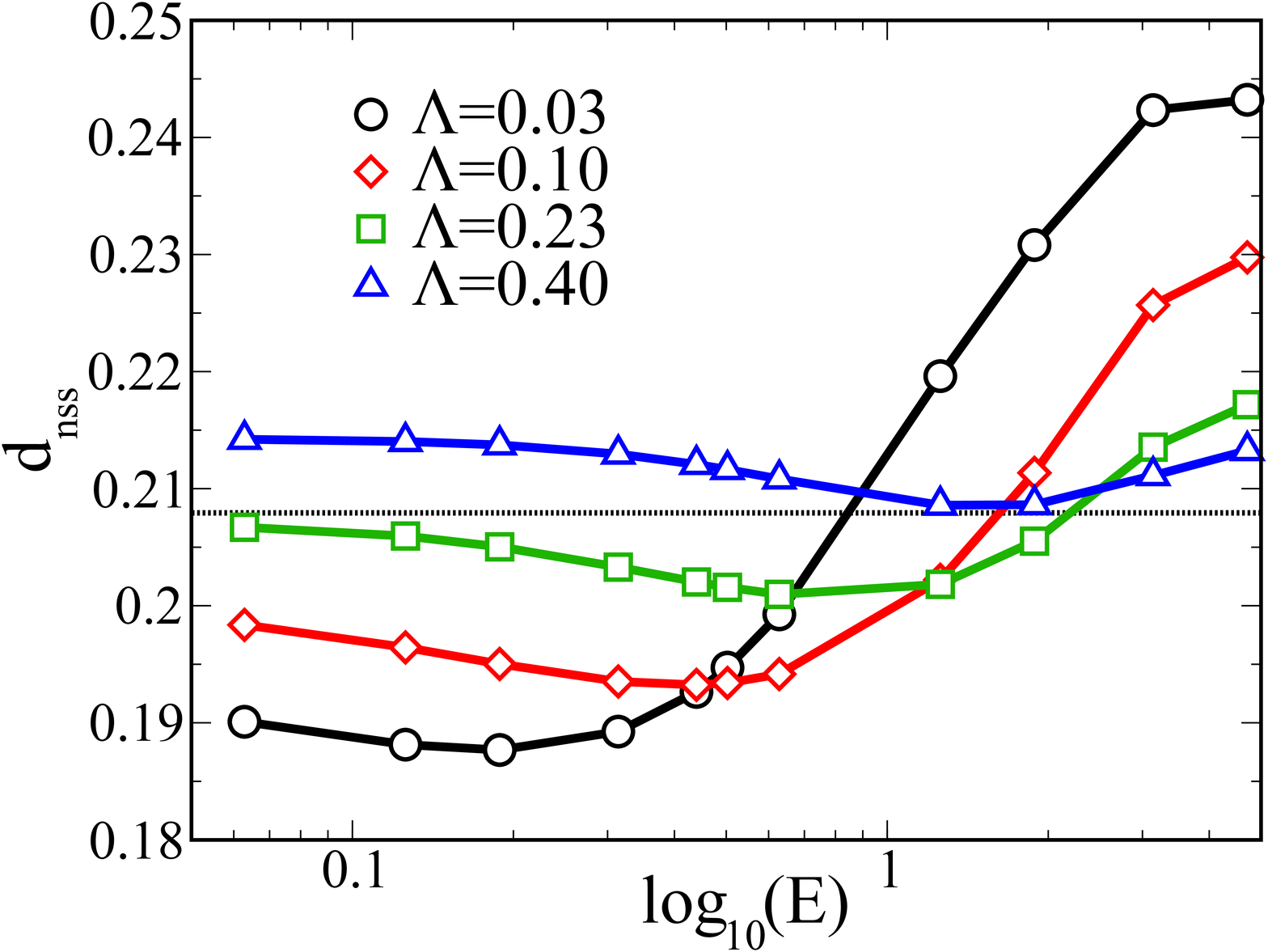}
\caption{(Color online) Double occupancy of the NSS $d_\mathrm{nss}$ as a function of the electric field for $U=6$ and increasing coupling to the thermostat $\Lambda$. The equilibrium solution for the same value of the interaction (dotted line) is reported for comparison.} 
\label{fig5}
\end{figure}

The behavior of the stationary double occupancy $d_\mathrm{nss}$ is plotted in panel (b) of \figu{fig5} as a function of $E$. $d_\mathrm{nss}$ is usually  different from the corresponding equilibrium value, reported in the figure for comparison.  Remarkably, the evolution of $d_\mathrm{nss}$ as a function of the electric field presents a minimum at a characteristic field strength. This behavior suggests that small applied fields, in conjunction with finite dissipation, increase the degree of electron localization in the NSS. Since the effective temperature of the NSS is a monotonically increasing function of the electric field, we can connect this result with an analog of the Pomeranchuk effect observed in equilibrium. In the latter case the more localized state is favored at finite temperature because of its larger spin entropy\cite{rmp}.
Upon increasing $\Lambda$,  $d_\mathrm{nss}$ rapidly reduces its range of variation, getting closer to the equilibrium value ($\Lambda=E=0$) for all values of $E$, implying that, in terms of correlation properties, the NSS becomes closer to the equilibrium state as the dissipation increases.

%

\begin{figure}
\centering
\includegraphics[width=0.49\textwidth]{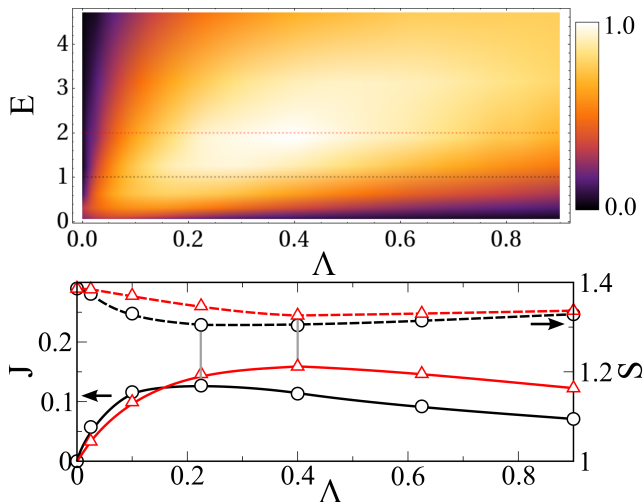}
\caption{(Color online) 
Top panel: phase-diagram of the thermostated non-equilibrium Hubbard model. Diagram is obtained from the normalized local current $J$ as a function of electric field strength $E$ and coupling to thermal bath $\Lambda$. Data for $U=4$. 
Horizontal lines indicates the cuts shown in the bottom panel.
Bottom panel: local current $\mathrm{J}$ and entropy $S$ behavior as a
function of $\Lambda$, for $E=1$ (circles), $E=2$ (triangles).}
\label{fig6}
\end{figure}

Finally, we investigated the dependence of the steady current on the coupling $\Lambda$ and the electric field. The results are cast in the ($E$-$\Lambda$) diagram shown in the top panel of \figu{fig6}. At the borders of the diagram we identify two small regions where the systems shows small or zero steady current. Near the $\Lambda=0$ axis the energy injected by the field largely overcomes that absorbed by the thermostat, and the electrons are constantly heated, thus leading to an incoherent motion with vanishing current. Conversely, near the $E=0$ axis, the large scattering introduced by strong coupling to thermostat reduces the linear conductivity and thus the corresponding steady current. In a wide intermediate region centered around the diagonal of the phase-diagram ($E/2\pi\simeq \Lambda$) we found largest values of the steady current. In this region the dissipation is sufficient to get rid of the extra energy pumped in by the field, but it is not too large to overcome the effect of the field driving a coherent current. To describe this effect more quantitatively we evaluated the "entropy" function $S=-2\sum_\k n_\k\ln(n_\k)$ which would coincide with the equilibrium entropy at $U=0$. The extension of entropy for non-equilibrium systems is a debated issue\cite{Polkovnikov08} which goes beyond the aim of this paper. Therefore we use $S$ simply as a tool to extract information about the approach to a stationary state and we do not interpret it as a physical entropy. The behavior of $S$ as a function of both $E$ or $\Lambda$ provides a way to relate the electronic states occupation with the values of the non-equilibrium stationary current. In the bottom panel of \figu{fig6} we compare the behavior of the current $J$ and that of $S$ as a function of increasing coupling $\Lambda$ for two fixed values of the current, corresponding to the horizontal cuts indicated in the diagram. The data point  out how, in the intermediate region, the maximal value of the electric current is achieved where $S$ is minimal.

\section{Conclusions}
Using DMFT in combination with a direct solution of the Kadanoff-Baym equations we 
investigated the non-equilibrium dynamics of the two-dimensional driven  Hubbard model
coupled to electronic reservoirs.
We reported that for a generic value of the field the coupling to an external bath is a 
necessary and, remarkably, also a sufficient condition to reach a non-stationary steady
state with a finite current. 
We characterized the properties of the NSS in terms of experimentally accessible quantities
and studied their dependence on the coupling to the thermostat,  
identifying the conditions to obtain a maximum of the steady current for a given field.
Our work provides a significant step towards a satisfactory description of non-equilibrium
solids, in which a certain degree of dissipation is always present. 
An explicit coupling to a thermostat is shown to be essential to obtain a description of a 
non-equilibrium stationary states with finite effective temperature, in contrast to a
modelling which neglects dissipation effects and that can only give rise to transient states
whose relevance for the physics of actual materials remains questionable.

\paragraph*{Acknowledgments.} 
A.A. thanks M.Fabrizio, M.Schir\`o and C.Aron for useful discussions. 
A.A. and G.K. have been supported by NSF-DMR-0906943. 
C.W. was supported by the Swiss Foundation for Science (SNFS).
M.C. and A.A. acknowledge financial support from the European Research Council under
FP7 Starting Independent Research Grant n.240524 ``SUPERBAD".
\bibliography{bibliografia}
\end{document}